\documentclass[lettersize,journal]{IEEEtran}

\usepackage{tikz}
\usepackage{amsmath}

\usepackage{filecontents}
\usepackage{cite}
\usepackage{amsmath,amssymb,amsfonts}
\usepackage{graphicx}
\usepackage{textcomp}
\usepackage{tabularx}
\usepackage{subcaption}
\usepackage{makecell}
\usepackage{url}
\usepackage[switch]{lineno}
\usepackage{listings}
\usepackage{multirow}
\usepackage{graphicx}
\usepackage{adjustbox}

\usepackage[margin=0.5in]{geometry}
\usepackage{textcomp}

\usepackage{listings}
\usepackage{tikz}
\def\checkmark{\tikz\fill[scale=0.4](0,.35) -- (.25,0) -- (1,.7) -- (.25,.15) -- cycle;} 
\colorlet{punct}{red!60!black}
\definecolor{background}{HTML}{EEEEEE}
\definecolor{delim}{RGB}{20,105,176}
\colorlet{numb}{magenta!60!black}
\graphicspath{ {./} }

\begin{document}

\title{ Containerization of a polyglot microservice application using
Docker and Kubernetes 
}

\author{Vamsi Krishna Yepuri,  Venkata Kalyan Polamarasetty, Shivani Donthi, Ajay Kumar Reddy Gondi \\
Department of Computer Science, Kent State University}



\maketitle

\begin{abstract}
  
This project investigates the benefits of containerization technology in modern software development and deployment. The study emphasizes the advantages of using Kubernetes and Docker in the development process, including the easy packaging and deployment of microservices, efficient resource utilization, faster startup times, and greater scalability and flexibility. The project concludes by proposing a study that involves creating a polyglot microservice application using Java, Python, and JavaScript, containerizing it with Docker, and deploying it in Kubernetes. The study aims to evaluate service discovery and auto-scaling in distributed mode and compare the performance metrics with virtual machines and containers. The results of this study can inform software development teams about the benefits of containerization in modern software development and deployment.

\end{abstract}

\begin{IEEEkeywords}
  Docker, Kubernetes, Containers,  Microservices, 
   Polyglot
  \end{IEEEkeywords}

\maketitle

\section{Introduction}
Containerization has transformed software development and deployment with Docker being one of the most widely adopted open-source platforms for building, shipping, and running distributed applications. However, as containerized applications become more complex, managing them at scale can be a challenge. Kubernetes, an open-source container orchestration system, provides a platform for managing and deploying containerized applications at scale. Its popularity is attributed to its ability to automate deployment, scaling, and management of containerized applications, with features like automatic load balancing, scaling, and self-healing. Trends in Kubernetes technology include the adoption of serverless computing, which enables developers to run their code in lightweight containers without managing infrastructure or container orchestration. Kubernetes also finds use in hybrid and edge computing, where computing resources are distributed across multiple devices and locations, demonstrating its versatility and adaptability to modern software development and deployment needs.

Public cloud providers have played a significant role in the adoption and advancement of container technology. They offer container services that allow developers to deploy, manage, and scale containerized applications, which have reduced the barrier of entry to container adoption, making them available to businesses of all sizes. These providers have also been instrumental in promoting container standards and enabling interoperability, contributing to the adoption of Kubernetes as the de facto standard for container orchestration. Standardization of container orchestration simplifies the deployment and management of containerized applications across different cloud providers and on-premises environments.

Containers provide a variety of benefits in microservices architecture, with one of the most significant being the ease of packaging and deploying microservices. Each microservice can be packaged as a container image, which abstracts it from the underlying infrastructure and makes it simpler to deploy and run the services across various environments without compatibility issues. For example, consider an e-commerce application based on microservices, with various services such as product catalogs, shopping carts, payment processing, and order management. Each of these services can be containerized, making it easy to deploy and update them independently. This enables developers to add new features or modify existing ones without impacting other services.

Additionally, containers provide a high degree of isolation between microservices, minimizing conflicts and dependencies between them. By packaging each service as a container, they can each run their own instance of the required database, without impacting the other services in the system. As a result, containers provide greater agility, scalability, and consistency in a microservices project, allowing developers to focus on developing and updating microservices rather than the infrastructure and deployment processes.

The use of virtual machines (VMs) can present challenges when it comes to running applications, as VMs require their own operating system and consume significant amounts of CPU, memory, and storage resources, which can result in slower application performance and increased complexity. This can be solved through the use of container technology, which allows multiple containers to share the same operating system kernel, leading to better resource utilization and faster startup times for applications, and ultimately greater scalability, flexibility, and cost efficiency.

For instance, consider a data analytics application that needs to process a large amount of data quickly and efficiently. If the application runs on a virtual machine, it may suffer from slower performance due to the overhead and complexity of running a full operating system. By using container technology, however, developers can run multiple lightweight containers on the same host operating system, resulting in faster performance, lower resource usage, and greater scalability for the application, allowing for faster data processing and analysis.

Achieving container capabilities with virtual machines is challenging due to the fundamental differences between the two technologies. Containers are lightweight and efficient, sharing the same operating system kernel as the host machine, while virtual machines emulate an entire hardware environment, including a separate operating system, which can lead to higher resource usage and greater complexity. Additionally, virtual machines lack portability and flexibility as each one requires a specific configuration of hardware and operating system, making it difficult to move or replicate them across different environments. Containers, on the other hand, can be easily packaged and deployed on any host system that supports the same container runtime. Although virtual machines have their own strengths and use cases, containers are preferred for most modern applications due to their greater efficiency, flexibility, and portability, and because they can start up much more quickly, making them better suited for modern application deployment and scaling.

The traditional monolithic application approach involves building a single, comprehensive app, which can be time-consuming to develop and maintain, making it challenging to achieve agility. To address these issues and achieve greater flexibility and ease of maintenance, the world is moving towards microservices. In the era of microservices, numerous independent modules are set up in traditional virtual machines, which can lead to conflicting libraries and make the process of provisioning, scaling, service discovery, load balancing, and deployment manual and time-consuming. To solve these problems, microservices can be packed in containers instead of virtual machines, which will help to address conflicts and create independence between each microservice. Workloads can be run in a scalable and distributed manner using Kubernetes, which can solve auto-scaling and service discovery issues. This project involves building a polyglot microservice~\cite{khine2019review} application from the ground up using Java, Python, and JavaScript, containerizing it with Docker, and deploying it in Kubernetes to examine service discovery and auto-scaling in distributed mode, while comparing the performance differences between the containerized technique and the traditional virtual machine strategy.

In this project, we aim to compare the performance of virtual machines and containers using several metrics. Firstly, we will examine the resource utilization of the microservice application between the two technologies, analyzing the CPU, memory, and storage usage to determine which technology is more efficient. Secondly, we will compare the startup time and scaling capabilities of virtual machines and containers to see which is faster and more scalable. Thirdly, we will examine the storage space utilized by both virtual machines and containers to see which technology uses storage more effectively. Finally, we will evaluate the portability of both virtual machines and containers and their ability to be moved across different environments. By comparing these performance metrics, we aim to determine which technology is more suitable for our microservice application and gain insights into the strengths and weaknesses of each technology.

\section{Related Work}
Microservices, Docker, Kubernetes, and polyglots are some of the major collectively used technologies in the field of software development. There are many works that can be found on these topics which demonstrate the purpose of a shift towards the docker and Kubernetes along with the challenges involved in the approaches. The importance of adopting a microservice-based architecture to achieve high availability for stateful applications in Kubernetes is explained in    ~\cite{vayghan2019microservice}. The challenges of managing stateful microservices in Kubernetes and also proposed a solution to improve their availability using a State Controller but did not mention the performance of Stateless polyglot micro service. In ~\cite{sharma2020docker}, presented a case study of a web application that consists of several microservices, each running in a separate container, and highlighted the  improved portability, scalability, and flexibility. There are some situations that demand the use of polyglots in order to achieve the desired solutions. ~\cite{zhang2019grit} discussed the approaches for achieving consistent distributed transactions across polyglot microservices that use multiple databases and addressed the challenges in distributed transactions in a polyglot environment. ~\cite{medel2016modelling} proposes a performance model to analyze resource management in Kubernetes clusters. The model uses queuing theory and stochastic processes to represent the resource utilization of nodes and pods. The proposed model is validated through experiments that compare its accuracy against the real system performance. The paper highlights the importance of having an accurate performance model to optimize resource allocation and improve the overall efficiency of Kubernetes clusters. Also in ~\cite{amaral2015performance} the evaluation of the performance of microservices architectures using containers, specifically Docker. The authors conduct experiments to compare the performance of a monolithic application with a microservices-based application, both deployed on Docker. The experiments involve measuring the response time and throughput of the applications under various workloads. The results of the experiments demonstrate that the microservices architecture outperforms the monolithic architecture in terms of response time and scalability. 
Overall, the use of microservices, Docker, Kubernetes, and polyglots in software development has become increasingly popular in recent years due to their benefits in terms of portability, scalability, and flexibility. However, there are also challenges involved in using these technologies, such as managing stateful microservices in Kubernetes and achieving consistent distributed transactions across polyglot microservices. To address these challenges, researchers have proposed various solutions and approaches, such as using a State Controller to improve the availability of stateful microservices in Kubernetes and developing a performance model to optimize resource allocation in Kubernetes clusters. In addition, studies have been conducted to evaluate the performance of microservices architectures using containers like Docker, and the results have shown that microservices architectures can outperform monolithic architectures in terms of response time and scalability. Overall, the adoption of microservices, Docker, Kubernetes, and polyglots in software development is likely to continue to grow as more organizations seek to achieve the benefits these technologies provide while also addressing the challenges involved in their use.

From \emph{TABLE I: Related Work}, we can see the different features implemented by the referenced papers and the features that are implemented in this project.
Our proposed solution will be dealing with a stateless polyglot micro service, containerized used docker, and Kubernetes. We deploy our application and analyze the improved performance using Kubernetes.

\begin{table}[h]
  \centering{
  \begin{tabular}{|l|l|l|l|l|l|} 
  \hline
  Reference & Microservices & Stateless & Container & Kubernetes & Polyglot  \\ \hline
  ~\cite{vayghan2019microservice} & \checkmark & X & \checkmark & \checkmark  & X \\ \hline 
   ~\cite{sharma2020docker} & X & \checkmark & \checkmark & X  & X \\ \hline
  ~\cite{zhang2019grit} & X & X & \checkmark & X  & X \\ \hline 
   ~\cite{medel2016modelling} & \checkmark & \checkmark & \checkmark & \checkmark  & X \\ \hline
  ~\cite{amaral2015performance} & \checkmark & X & \checkmark & X  & \checkmark \\ \hline
Our Project & \checkmark & \checkmark & \checkmark & \checkmark  & \checkmark \\ \hline
     
  \end{tabular}
  }
\caption{Related Work}
\end{table}
\section{Project background}
The trend towards microservices architecture has become more and more popular in recent years, as it allows organizations to develop, deploy, and manage large, complex applications in a more agile and efficient manner. However, one of the challenges of implementing a microservices architecture is managing the deployment and scaling of individual services in a way that is both effective and efficient. This is where containerization and container orchestration tools like Docker and Kubernetes come into play. In this project, we are going to show the analysis of performance improvement with the use of container orchestration tools.

\subsection{ Microservice vs Monolithic architecture}
Monolithic and microservice architectures are two different approaches to designing and building software systems. A monolithic architecture is a traditional approach where an application is built as a single unit, while a microservice architecture is an approach where an application is broken down into smaller independent services. \emph{Fig. 1: Monolithic vs Microservices} shows the difference between monolithic and microservice-based architecture. The monolithic approach is easier to develop and deploy but can become difficult to maintain and scale as the application grows. On the other hand, the microservice approach offers greater flexibility, scalability, and resilience but requires coordination between different teams. The choice of architecture depends on the specific needs and requirements of the application, with monolithic being better suited for smaller applications and microservices for larger, complex applications.

\begin{figure}
\centering
\includegraphics[width=9cm, height=5.5cm]{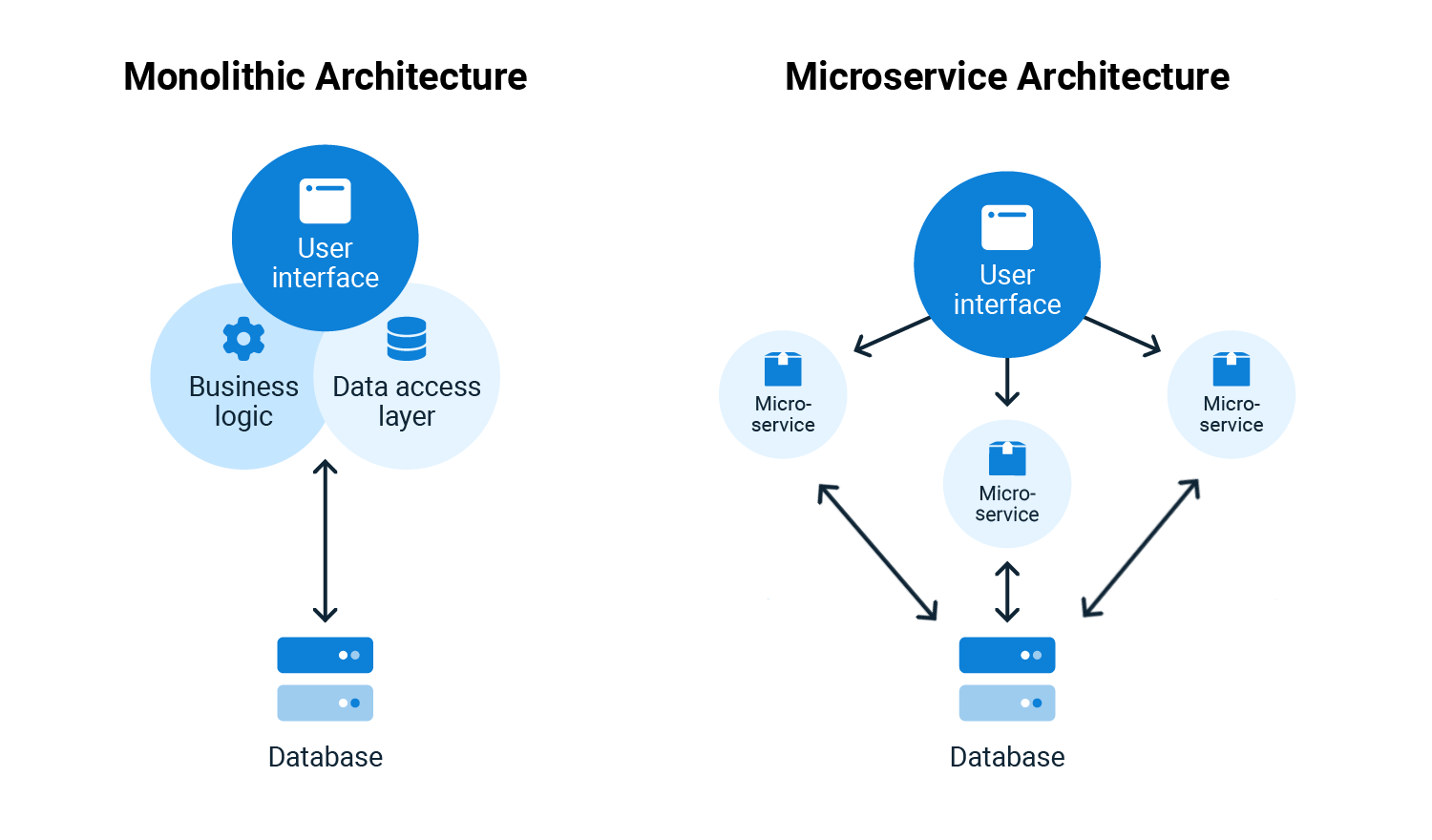}
\caption{Monolithic vs Microservices}
\end{figure}

\cite{escobar2016towards}explains the process of transforming monolithic applications into microservices and provides a framework for understanding the migration process. It discusses the benefits and challenges of adopting a microservices architecture.

\subsection{ Polyglot microservices}
Polyglot microservices refer to a software architecture that involves using multiple programming languages and technologies to build independent, small-sized services that work together to accomplish a larger goal. \emph{Fig.2 : Polyglot microservices} represents polyglot microservices. This approach allows developers to choose the best language and tools for each service, rather than being limited to a single technology stack. Polyglot microservices are becoming increasingly popular because they offer several benefits, such as increased flexibility, improved scalability, and better fault tolerance. They can be designed to communicate with each other through APIs, message queues, or other protocols, which enables them to work together as a cohesive system.
While there are challenges to building and managing polyglot microservices, such as the need to manage multiple programming languages and tools, these can be overcome with the right approach and tools. As a result, many organizations are adopting this approach to build more resilient and adaptable software systems.

\begin{figure}
\centering
\includegraphics[width=5.5cm, height=6cm]{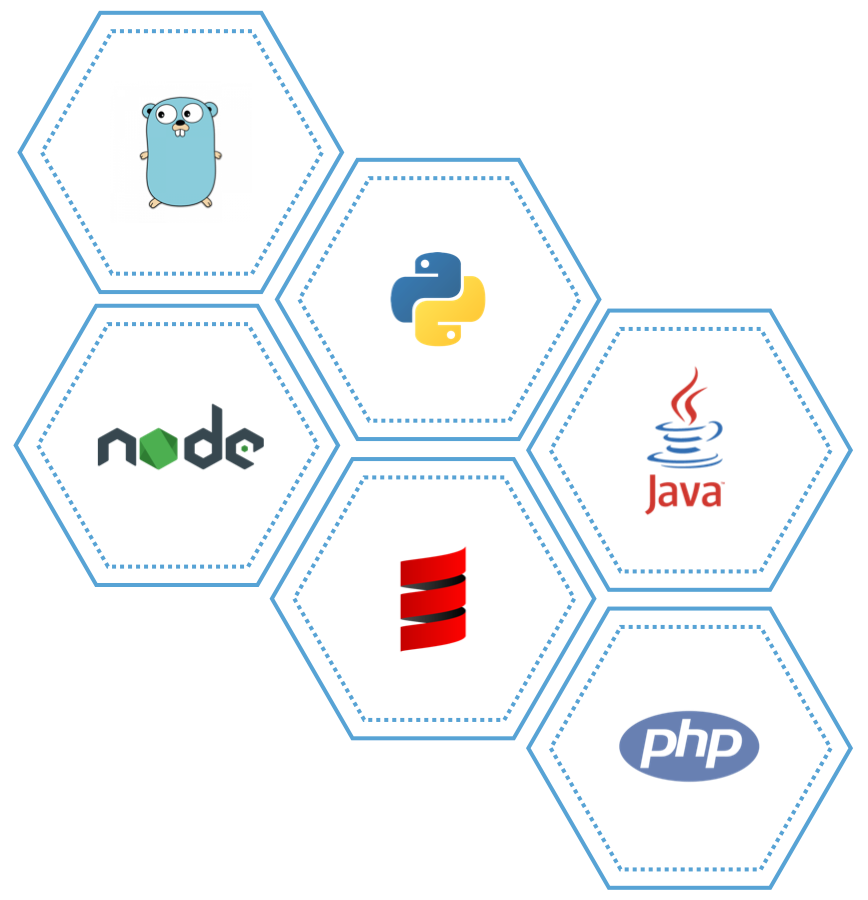}
\caption{Polyglot Microservices}
\end{figure}

A polyglot microservices architecture can provide significant advantages for monitoring~\cite{rak2011cloud}, 
load balancing~\cite{9676704}, security~\cite{zdun2023microservice} and QoS~\cite{9502607} by allowing you to 
choose the best technology for each microservice, improve resilience, make monitoring and troubleshooting easier, distribute the workload, and optimize each microservice for specific tasks or functions.
A polyglot microservices architecture can also take advantage of traffic engineering!\cite{robin2022p4te} capabilities in data center networks to 
optimize the performance of microservices~\cite{cortellessa2022model} and improve overall application performance. 
By leveraging intelligent routing, traffic shaping, load balancing, network performance monitoring, and scalability, 
a polyglot microservices architecture can ensure that traffic is routed to the most optimal endpoint, manage the flow of 
traffic between microservices, distribute traffic evenly across multiple endpoints, monitor network performance, 
and accommodate changes in traffic volume and application usage. All of these capabilities can help to reduce latency, 
prevent congestion, allocate resources efficiently, identify and resolve issues quickly, and ultimately, improve application performance and user experience.

\subsection{ Container}
A container is a lightweight, standalone, and executable package of software that includes everything needed to run an application, such as code, libraries, and system tools. Containers provide a consistent run time environment, ensuring that the application can run reliably regardless of the underlying system's configuration. \emph{Fig. 3: Virtual Machine vs Container} shows the difference between a virtual machine and a container.
~\cite{siddiq2014comprehen} is a comprehensive report on the use of container technology in modern software development. It covers various aspects of container technology, including its history, benefits, challenges, and future trends. The goal of containerization is to package each service and its dependencies into a lightweight, portable container that can run on any machine or cloud environment. 

\begin{figure}
\centering
\includegraphics[width=9cm, height=5cm]{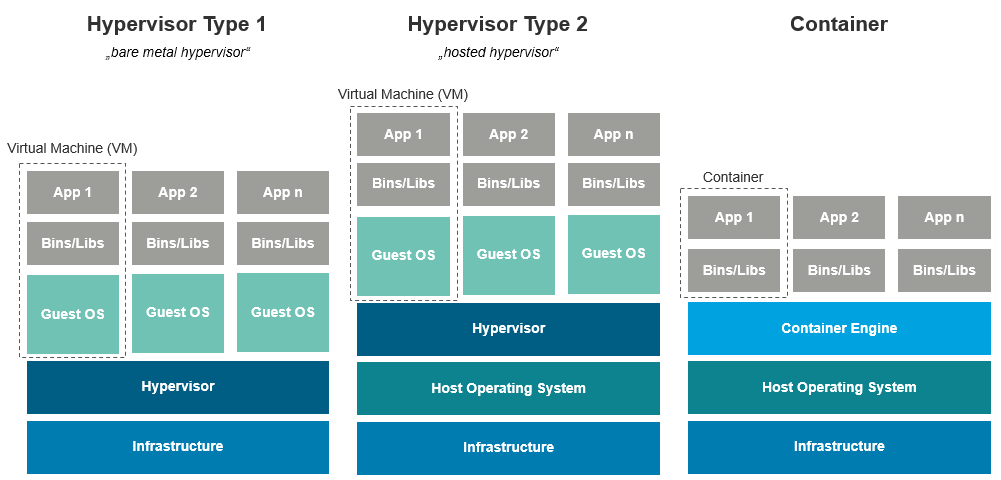}
\caption{Virtual Machine vs Container}
\end{figure}

Docker is widely used as a containerization platform because it offers a reliable and reproducible setting for running applications. Docker makes it simple to create, deploy, and manage images, which makes it easier to develop and test applications across different environments. A Docker container is a self-contained software package that operates independently of other containers and the host system. It has its own network, file system, and resources, making it perfect for deploying applications with specific requirements. Docker images serve as snapshots of an application's file system, and they can be created using a Dockerfile script. Images can be utilized to start one or more containers. Docker containers have advantages such as portability, consistency, isolation, resource efficiency, and scalability, while their disadvantages include complexity, a learning curve, security concerns, dependency management challenges, and limited access to hardware resources.

\subsection{Kubernetes and Docker}
Kubernetes is an open-source platform for container orchestration, which automates the deployment, scaling, and management of containerized applications.\emph{Fig. 4: Docker vs Kubernetes} shows the difference between docker and Kubernetes and how Kubernetes is used in a distributed environment and orchestrating the docker containers. It provides a framework for deploying, scaling, and managing multiple containers across a cluster of servers, allowing organizations to run and manage their microservices at scale.
~\cite{kho2018auto} explains how Docker and Kubernetes can be used to automate the scaling of a defense application in a cloud environment. It highlights the benefits of these tools, such as improved security and reliability, and provides a detailed deployment process.

\begin{figure}
\centering
\includegraphics[width=9cm, height=4.5cm]{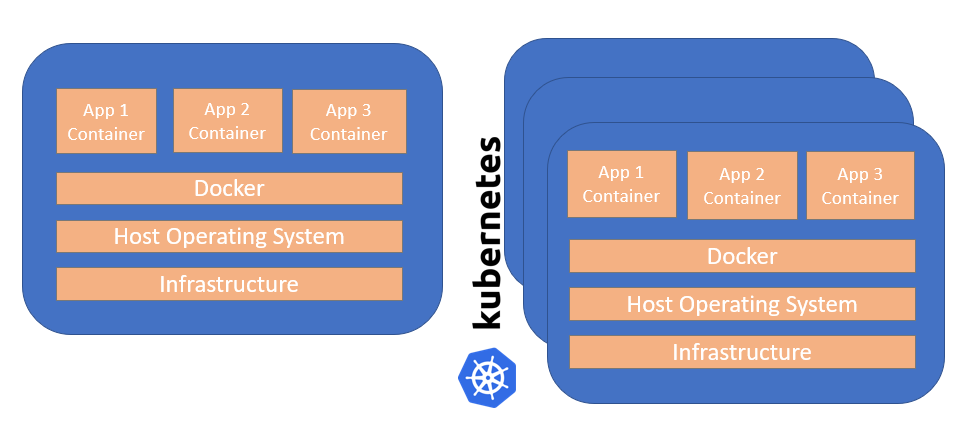}
\caption{Docker vs Kubernetes}
\end{figure}

In the case of a polyglot microservices application, where multiple programming languages and frameworks are used to develop individual services, containerization, and container orchestration can be particularly valuable. Docker and Kubernetes enable organizations to easily package and deploy services written in different languages or frameworks while ensuring that each service runs in a consistent and isolated environment. This makes it easier to manage dependencies, test and deploy services, and scale applications as needed.

Overall, the containerization of a polyglot microservices application using Docker and Kubernetes can help organizations achieve greater agility, scalability, and reliability in their software development and deployment processes. It can also help reduce the complexity of managing and scaling multiple services, which can be particularly challenging in a polyglot environment.
\section{Infrastructure}
The infrastructure required for containerization using Docker and Kubernetes typically includes :

a. physical or virtual machines,

b. storage systems, and 

c. networking components. 

Physical or virtual machines are needed to host containerized applications. These machines should have sufficient CPU, memory, and storage to run the application and the containers. Storage systems are necessary to store the container images, as well as any data or files that the application needs to access. The storage should be accessible to all the machines in the infrastructure. Networking components such as load balancers~\cite{9676704} and firewalls are also required to ensure that traffic can be routed to the appropriate containers and that the network is secure.
We also need a cloud engine, for that we are using GCP. Google Cloud Platform (GCP) is a cloud provider that offers a range of services for hosting and managing applications, data, and infrastructure in the cloud. GCP provides tools and services such as Google Kubernetes Engine (GKE), Google Container Registry (GCR), and Cloud Storage that are well-suited for containerization using Docker and Kubernetes. Additionally, GCP provides tools for monitoring and managing containerized applications, such as Stackdriver Monitoring and Logging, and Cloud Trace for application tracing.

There are examples provided in ~\cite{gupta2020deploy} describe the steps to deploy a web application using Google Cloud Platform, which involves creating a new project, enabling necessary APIs, choosing a deployment method, creating an instance or app, configuring it, and deploying the application. Once deployed, the application can be managed and monitored through the Google Cloud Platform.

Also, the GCP services used were VPC networks (Virtual Private Cloud Networks) and Compute Engine. VPC networks allow for the creation of private virtual networks within GCP, while Compute Engine provides virtual machines with configurable amounts of CPU, memory, and storage for running Kubernetes and deploying applications and services.
\begin{figure}
    \centering
    \includegraphics[width=10cm, height=5.8cm]{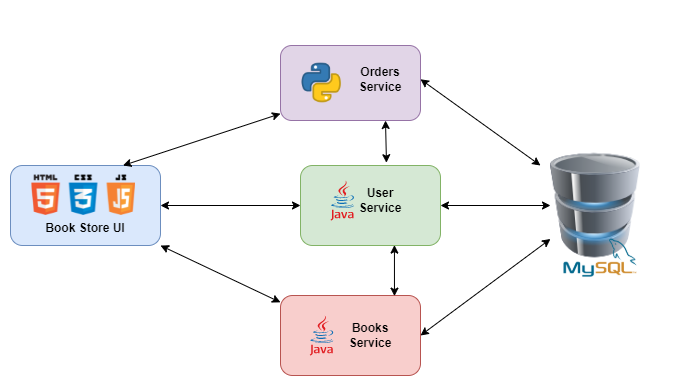}
    \caption{Bookstore application Architecture}
    \label{fig:my_label}
\end{figure}
\section{System Architecture}
To carry out a test of Docker and Kubernetes with an application in our scenario, it is necessary to have a polyglot microservice application that is designed with various programming languages. The bookstore application comes into picture here. The bookstore application has been designed to offer its customers a convenient way to browse and purchase books online as a polyglot microservice. Four microservices are being used to fulfill the result of the application as seen in Fig.5 which are developed in multiple programming languages and using scripts. They are the user service, orders service, books service, and the bookstore UI service. The data is stored in the central repository of a database. The microservices are all connected to one another to communicate between them using REST APIs.
\begin{figure}
    \centering
    \includegraphics[width=9cm, height=5.8cm]{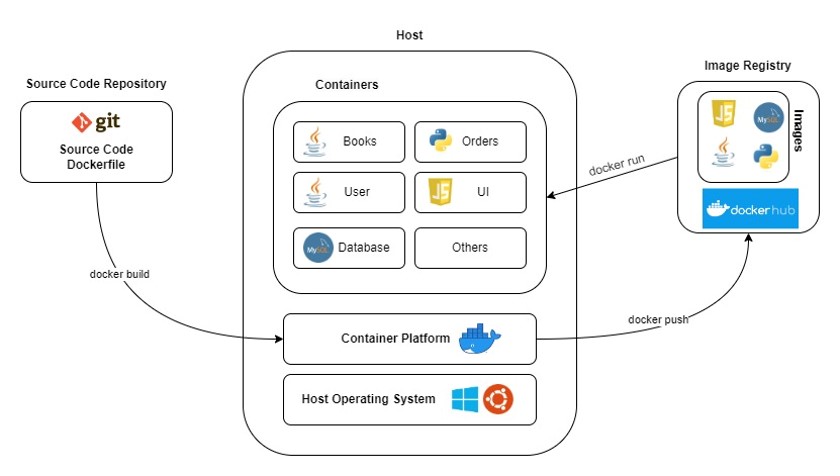}
    \caption{Architecture of Bookstore application on single-node Docker system}
    \label{fig:my_label}
\end{figure}
Let us suppose that the bookstore application is deployed on a single-node Docker system as shown in Fig.6. Docker containers are lightweight and standalone virtual machines that can run on any computer. The Docker engine effectively manages these containers and ensures that they have access to the necessary resources, such as memory and CPU, while keeping them isolated from each other. By running multiple containers on the same machine, each with its own set of dependencies and configurations, applications can be easily deployed and managed without having to worry about conflicts between different software components. All the services with the source code and Dockerfile are built using the Docker container platform. Once the build is completed, the data is pushed to the central repository of the Docker hub.  This eliminates the need to download and configure software dependencies for each microservice individually, making deployment more efficient and less time-consuming. With the required software already pre-packaged into the images, it's simply a matter of pulling the necessary images and running them in the Docker container. To run the bookstore application, all the microservices required images that are saved on Docker Hub are set to run by using the docker run command. The containers are created once the docker run with the image name saved in the Docker hub command is given. The main reason of moving from the single-node Docker system to Kubernetes is that it can provide several benefits such as scalability, high availability, resource efficiency, service discovery for more complex applications with higher demands from the user. By distributing the workload across multiple nodes, Kubernetes can handle much larger workloads than a single-node Docker system, making it an ideal choice for applications with rapidly increasing traffic. The automatic detection and recovery features of Kubernetes can help maintain application availability, reduce downtime, and minimize the need for manual intervention.

The bookstore application can be deployed on Kubernetes. For the multiple nodes, we are using the Google Cloud platform - GCP as shown in Fig.7. The system starts with the creation of a VPC network, which is a virtual private network for Google Cloud resources. This network provides a private network space for the deployment of Kubernetes clusters. A Kubernetes cluster is then deployed on the VPC network with 1 master node and 2 slave nodes. This cluster consists of one or more nodes, where each node is a virtual machines (VMs). Each virtual machine runs a Kubernetes node, which is responsible for running pods and managing the resources of the virtual machine. Utilizing compute engine in GCP, a kubernetes master node is  deployed, which manages the Kubernetes API server, controller, scheduler, and the etcd database. The API server is responsible for receiving and processing requests from its clients, while etcd stores configuration data and metadata for the Kubernetes cluster. The controller sends the commands to the slave nodes.  For example, if the image is to be deployed in the slave node, the controller sends a command to the kubelet in the slave node to implement the process. The scheduler is used while the initial creation of the pods in the slave nodes to configure the placement of the pods in case of any downtime. The container run time is used to run the container when it is started. We will use Network File System (NFS) server to store the data stored in the databases as a central repository . Kubernetes clients can then interact with the API server to deploy and manage applications on the cluster. When an application is deployed, Kubernetes schedules the necessary containers onto the available nodes based on resource requirements and availability and also manages their life cycle, scaling them up or down as needed. To access the application, users connect to a load balancer that is deployed on the Kubernetes  as a service. The service distributes traffic across the containers running in the Kubernetes cluster.

\begin{figure}
    \centering
    \includegraphics[width=9.3cm, height=6.2cm]{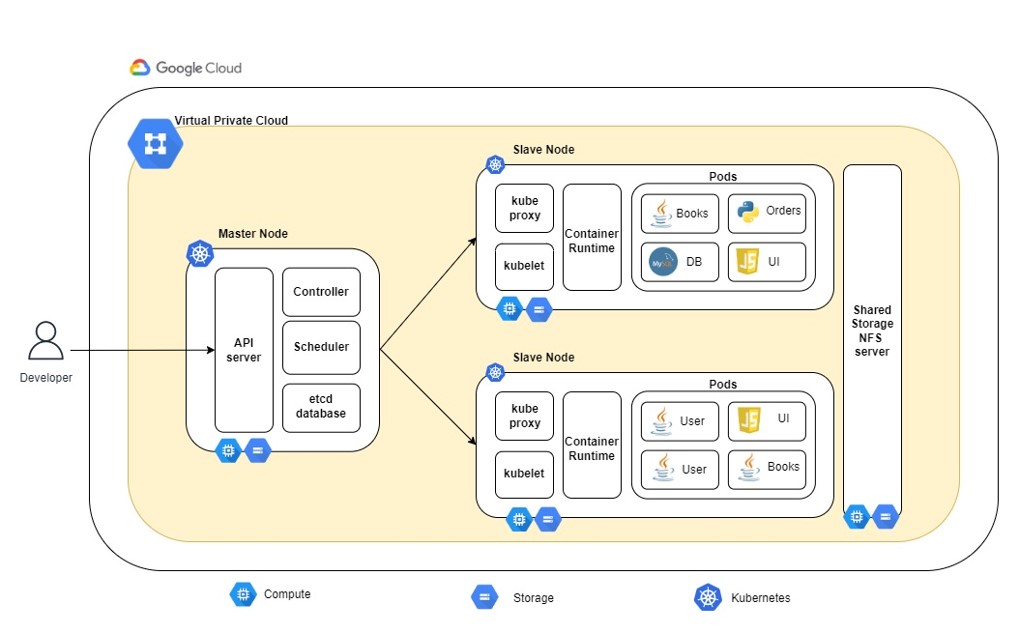}
    \caption{Architecture of Bookstore application on Kubernetes}
    \label{fig:my_label}
\end{figure}
\section{Implementation}
This project is implemented as the following four steps. To implement the containerization of polyglot microservice application , firstly we need to develop the microservices required for the application and then deploy the images of the microservices to the cloud system where kubernetes is configured. After deploying the microservices, we will compare the performance of the application on a virtual machine and a container.

\subsection{Development of polyglot microservices and containerizing then with Docker}
A bookstore application is developed to order books online. The customer is able to view the books, add books to the cart, order them, and view the order history and the status of the order. The administrator can also easily add new books to the application, including important details such as author information, price, and quantity available. To achieve its functionality, the bookstore application is comprised of four key microservices, which are the User Service, Orders Service, front-end Bookstore UI, and the Books Service. All of these microservices work together seamlessly to ensure optimal performance of the application. The data is stored in a centralized database, specifically MySQL, which has been carefully chosen for its reliability and performance. The Bookstore UI is developed using HTML, CSS, and JavaScript, providing a user-friendly and intuitive interface for customers and administrators. Depending on the user's role, the User Service ensures that the appropriate functionality is available to them. This  service has been developed using Java programming language. The Orders Service is responsible for managing the user's cart details, order history, and the delivery status of each order which has been developed using Python programming language and has been seamlessly integrated with the other microservices. The Books Service, developed in Java programming language, is responsible for managing information about the books available in the application.  The application has been developed with great attention to detail, ensuring that every aspect of the process, from browsing books to order fulfillment, is seamless and intuitively developed as polyglot microservices. Once the application was developed, we converted it into Docker images using Dockerfile and then pushed those images to Docker Hub. Next, we set up a single node Docker machine and pulled all the Docker images from Docker Hub onto this machine. Finally, we ran all the containers on this single node Docker machine, allowing us to test our application using Docker.

\subsection{Setting up Kubernetes on cloud}
For the deployment of the bookstore application in a large scale, the Kubernetes setup is done on the Google Cloud Platform (GCP). A Virtual Private Cloud (VPC) network is auto-configured in the Google Cloud Platform for the intercommunication of the applications and network discovery. As part of the project, a Kubernetes cluster of three nodes is created with the configuration of 4vCPU and 8GB of RAM. We are implementing the cluster has one master and two slave nodes. The operating system deployed on the machines is Ubuntu 20.04 LTS amd64 focal image built on 2023-03-02. We are using containerd as the container runtime. We have created one more node with the configuration of 4vCPU, 8GB of RAM, and 50GB SSD with Ubuntu 20.04 LTS amd64 focal image built on 2023-03-02 as an operating system. This machine is used as the Network File System - NFS server for centralized storage. In a Kubernetes environment, an NFS server can be used as a persistent storage solution that allows containers to access and manipulate data stored on a networked file system. The Kubernetes cluster is configured to automatically provision and manage persistent volumes (PVs) that are backed by the NFS server, which can be used by applications running in pods. When a pod requests access to the storage, it would create a PersistentVolumeClaim (PVC) that specifies the storage requirements, which Kubernetes would use to dynamically provision a PV that is backed by the NFS server. Once the PV is provisioned, it can be mounted as a volume in the pod's container, allowing the application to read and write data to the shared storage. This provides a scalable and reliable way to store data that can be accessed from multiple pods or nodes in a Kubernetes cluster.
\subsection{Deployment, scaling and load balancing of microservices on K8s}
We have created the manifest files required for the deployment of the bookstore application in this milestone. In Kubernetes, deployment and service files are two types of manifest files used to define and manage applications running in a cluster. For each one of the microservices, a deployment file is created as a deployment object in Kubernetes which in turn runs the application container inside a pod. The deployment file typically includes specifications for the container image, the number of replicas to create, and any necessary environment variables, volumes, and networking configuration. When the deployment file is applied to the cluster, Kubernetes will create the specified number of replicas and manage their life cycle, ensuring that the desired number of replicas is always running and up-to-date. This is responsible for scaling each microservice and auto-healing the pods. Service files for all the microservices are created as they are used to define a Kubernetes Service object, which provides a stable IP address and DNS name for accessing a set of pods running in the cluster. The service file typically includes specifications for the type of service, the selector used to target the pods, and any necessary networking configuration. When the service file is applied to the cluster, Kubernetes will create the Service object and ensure that it is available for other objects in the cluster to use for accessing the pods. By default, the Round robin scheduling algorithm is used for diverting the traffic to the pods. The MySQL database is configured as the persistent volume as the data stored in the MySQL database is persisted across pod restarts or node failures, ensuring that the data is always available. It also makes sure to use a centralized storage solution, such as an NFS server or cloud-based storage service, for storing the data. This can help ensure that the data is stored securely and centrally, and can be easily accessed by other pods or services in the Kubernetes cluster. We ensured that all four microservices of the book application were running as pods in the Kubernetes cluster and that they were able to communicate with each other and the database pods. Next, we exposed the application to the internet by creating a service. This allowed users to access the application using any IP of the node in the cluster.After completing these steps, the book application was successfully up and running, and users could access it over the internet using any of the IP of the node in the cluster.

\subsection{Performance comparison of application on VM and containers}
Performance evaluation for containers versus VMs is an important step in determining the most appropriate virtualization technology for a given application or workload.In this evaluation, the key metrics considered are

1. Boot Time 

2. Server Footprint

3. Performance metrics

For the evaluation of performance of containerized services and services running on VM , both the systems are tested in the same environment and allocated equal resources of 2 core CPU with 2GB RAM to achieve the best possible results.

\subsubsection{Boot Time}
Boot Time refers to the amount of time it takes for a server or system to start up and become fully operational. It's an important parameter to consider, especially in situations where high availability and recover time are critical, as longer boot times can result in longer down times and potentially impact user experience. Factors that can impact boot time include the hardware configuration, software stack, and network connectivity. In our scenario, applications are created as services inside VM and container, which means the application will be started after the system boots. Here we considering the boot times of both VM and a container.

a. Boot time of application running on a VM :  To calculate the boot time of the server and application to come up, the timestamp difference between the VM powered-on  and the application serving timestamp is calculated.

b. Boot time of application running on a container : To calculate the boot time of the application running on a container,the timestamp difference between the  time at which existing pod is killed and the next pod serving is calculated.

\begin{figure}[!h]
    \centering
    \includegraphics[width=9.3cm, height=4.8cm]{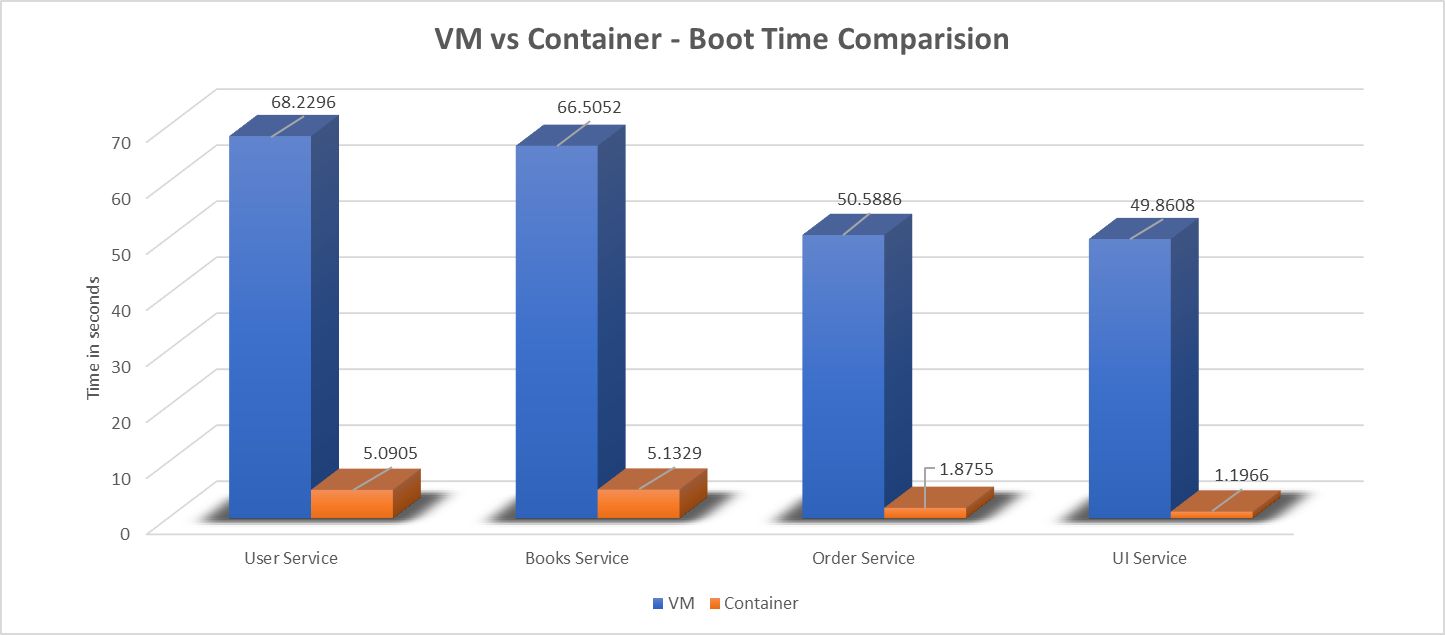}
    \caption{VM vs Container - Boot Time Comparison}
\end{figure}

 \emph{Fig. 8: VM vs Container - Boot Time Comparison} shows the comparison of boot time on VM and container and it is clearly evident that the boot time has been greatly improved using the containerized services.

\subsubsection{Server Footprint}
Server footprint refers the amount of physical space a server takes up. A smaller server footprint can be desirable in situations where space is limited. 

\begin{figure}[!h]
    \centering
    \includegraphics[width=9.3cm, height=4.8cm]{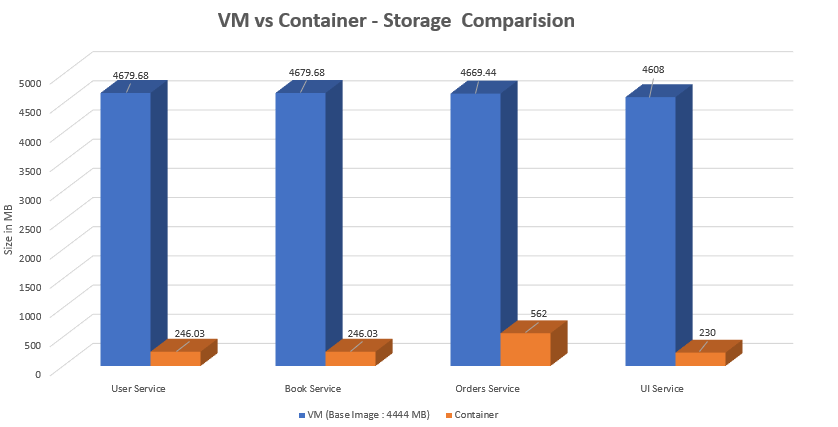}
    \caption{VM vs Container - Server foot Print Comparison}
\end{figure}

\emph{Fig: 9 VM vs Container - Server foot Print Comparison} depicts the tremendous decrease in the server footprint with the containers compared to that of VM's.

\subsubsection{Performance Metrics}
Performance Metrics refers to various metrics that are used to measure the performance of a server or system, such as cpu utilization, memory usage  and response time. These metrics are important to consider in order to ensure that the system is meeting the needs of its users and is performing optimally. Factors that can impact performance metrics include the hardware configuration, software stack, network connectivity, and workload characteristics. We have tested performance metrics for both VM and containers. We observed the performance metrics under two states. One is idle state, when servers are not handling any traffic. Another state is load state, where we simulate the load on the application. In this two different states, we observed the performance metrics for both the containers and VMs.

In load state, i.e to simulate the load on the application, we used JMeter tool where we simulate a load of 1000 requests in 1 second ramp up period with 10 iterations. So totally 10000 requests will be imposed on the application.  The observation of the VMs and containers during idle state and load state is carried out using Prometheus monitoring tool. To visualize these metrics, Grafana is used on top of prometheus to create dashboards for observing the performance. Each vm and container were allocated with the limits of 2 core CPU and 2 GB memory.  In the following, we compare the performance metrics for our polyglot microservices in different states (idle and load) in both VM and containers.

\paragraph { Orders - Python Service}
The orders microservice was written in python for the management of orders in our application. 

\begin{figure}[!h]
    \centering
    \includegraphics[width=9.3cm, height=4.8cm]{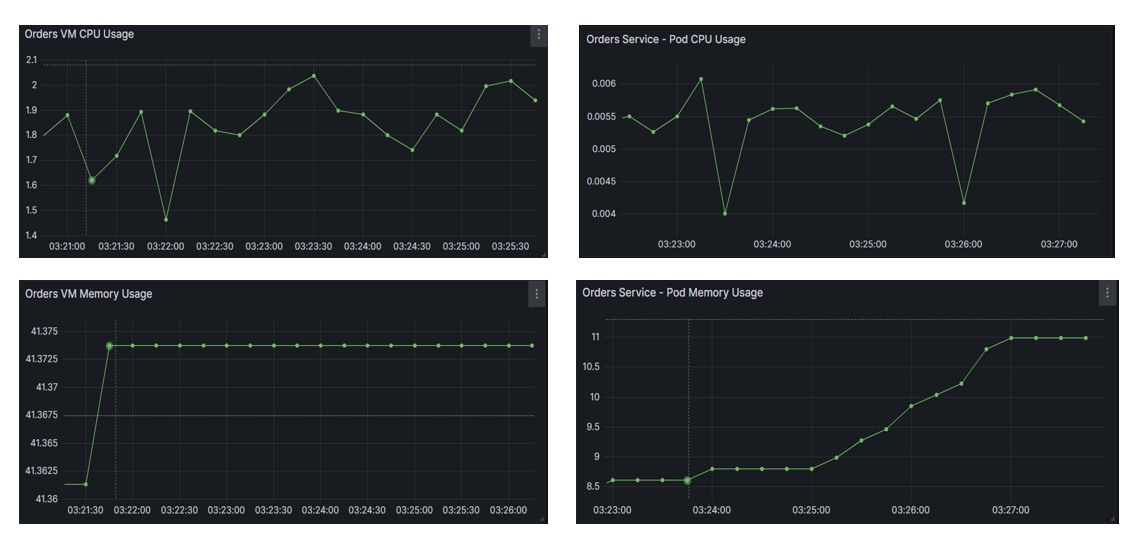}
    \caption{Orders Python Service - Idle state Comparison}
\end{figure}
\begin{figure}[!h]
    \centering
    \includegraphics[width=9.3cm, height=4.8cm]{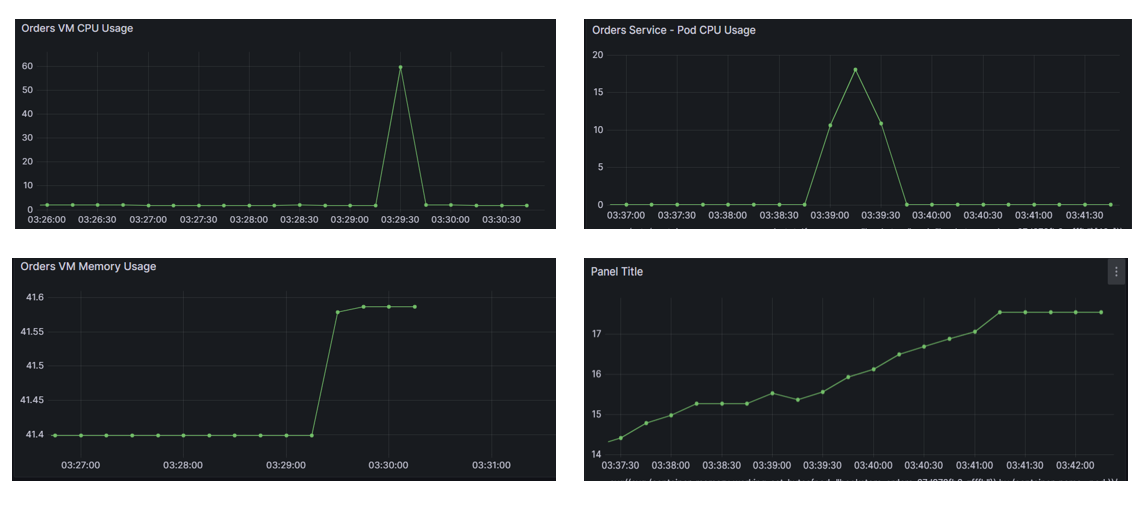}
    \caption{Orders Python Service - Load state Comparison}
\end{figure}
\begin{figure}[!h]
    \centering
    \includegraphics[width=9.3cm, height=3cm]{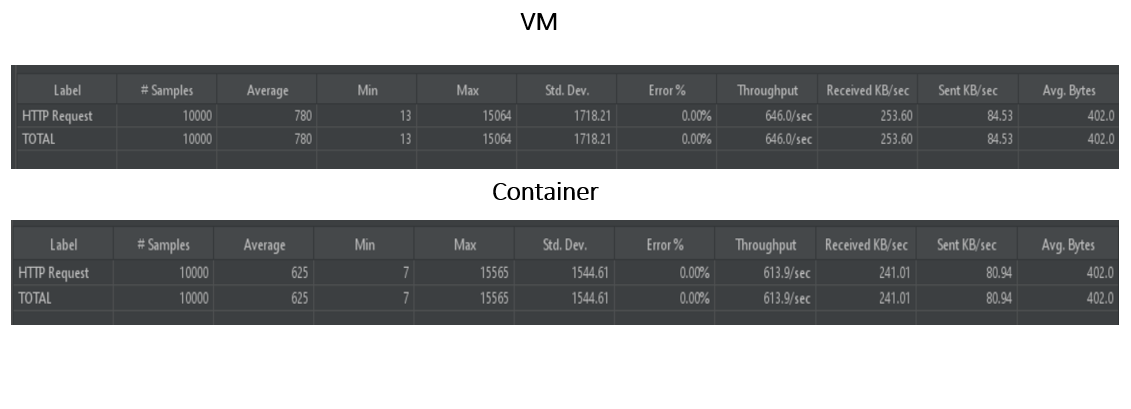}
    \caption{Orders Python Service - Response Time Comparison}
\end{figure}
 \emph{Fig: 10 Orders Python Service - Idle state Comparison} shows the comparison of the CPU utilization and memory consumption by the application on VM and container in the idle state. When in the idle state the CPU usage by the VM fluctuated between 1.4\% and 2.1\%. Whereas in the container, it fluctuated between 0.004\% and around 0.006\%. The Memory usage by the order service in VM started around 41.36\% and increased to around 41.3725\% and stayed constant. Whereas in the container, it started around 8.5\% and increased up to 11\% and stayed constant for a while. \emph{Fig: 11 Orders Python Service - Load state Comparison} shows the comparison of the CPU utilization and memory consumption by the application on VM and container in the load state. When the load was applied on the orders service running on the VM, we can see a significant spike in the CPU usage that ranged around 60\% for a while, whereas in the container we can also see a significant spike ranging around 20\%. The Memory usage by the order service in VM started around 41.4\% and increased to around 41.6\%, whereas in the container, it started around 14\% and increased up to 18\% and stayed there for a while. By the above comparisons of CPU usage and memory usage by the orders service running in VM and container, both in idle and load state, we can clearly state that the orders service consumes less resources in container. \emph{Fig: 12 Orders Python Service - Response Time Comparison}s hows the comparison of the response times of the orders service when running on VM and a container. While running the application in both VM and container, 10000 requests were sent to check the response times the average time was about 780ms on VM, whereas while running on the container the average response time was 625ms. By the above comparison, we can clearly see that the service performs well in a container.

\paragraph{ UI - JavaScript Service}

The UI is handled by a microservice was written in HTML and JavaScript to carryout the user interaction with application. 
\begin{figure}[!h]
    \centering
    \includegraphics[width=9.3cm, height=4.8cm]{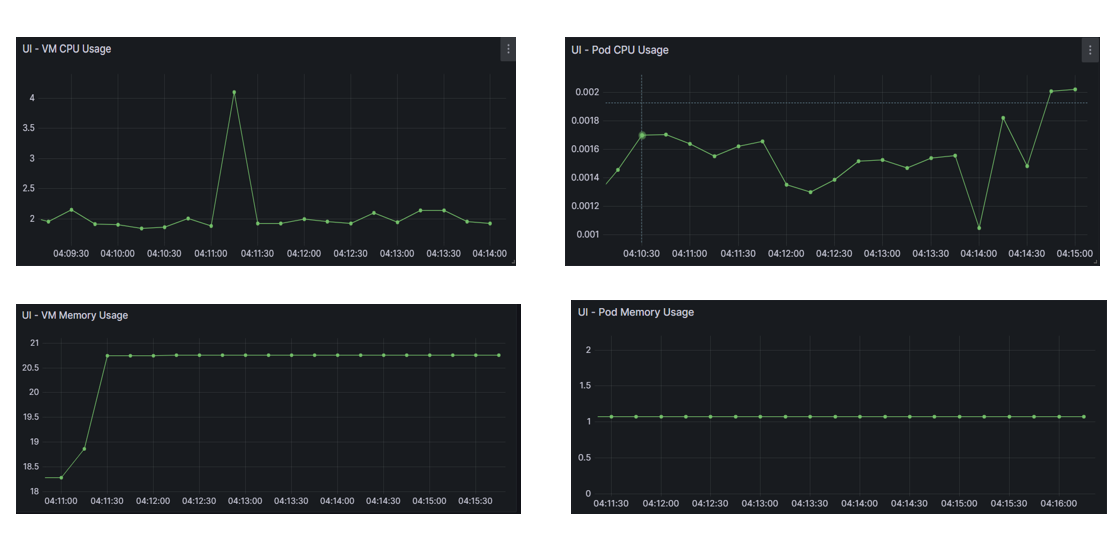}
    \caption{UI JavaScript Service - Idle state Comparison}
\end{figure} 
\begin{figure}[!h]
    \centering
    \includegraphics[width=9.3cm, height=4.8cm]{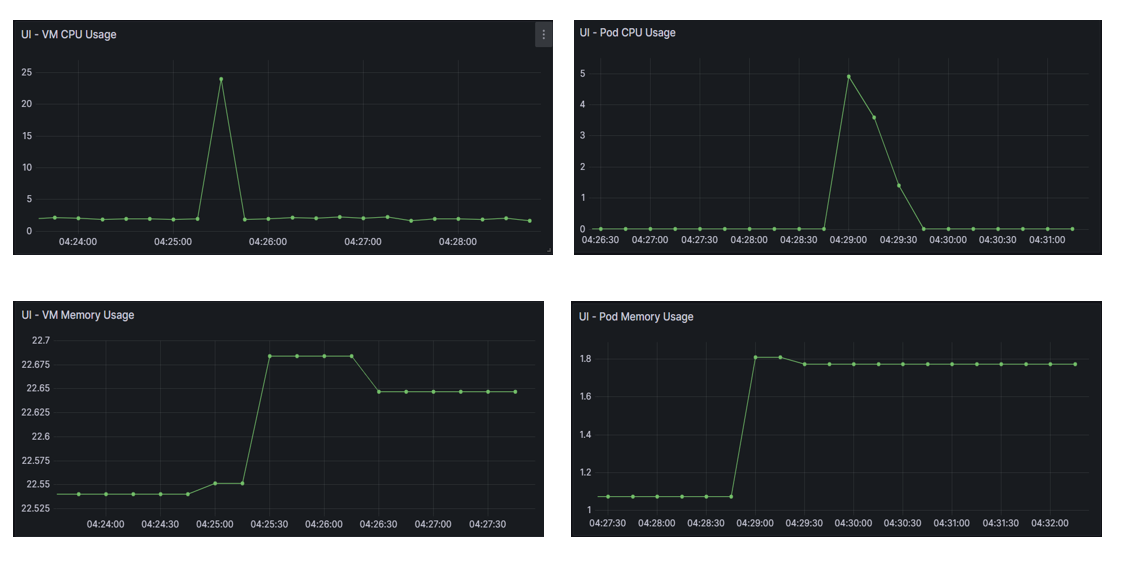}
    \caption{UI JavaScript Service - Load state Comparison}
\end{figure}
\begin{figure}[!h]
    \centering
    \includegraphics[width=9.3cm, height=3cm]{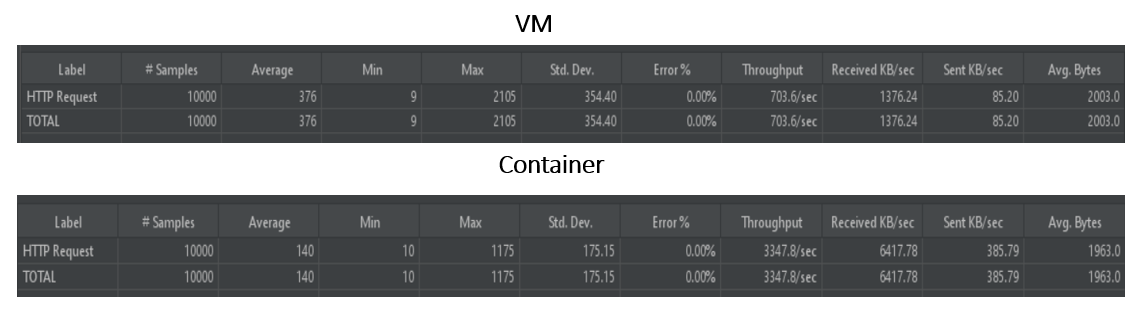}
    \caption{UI JavaScript Service - Response Time Comparison}
\end{figure}
\emph{Fig: 13 UI JS Service - Idle state Comparison} shows the comparison of the CPU utilization and memory consumption by the application on VM and container in the idle state. When in the idle state the CPU usage by the VM fluctuated between 2\% and around 4\%, whereas in the container, it fluctuated between 0.001\% and around 0.002\%. The Memory usage by the UI service in VM started around 18\% and increased to around 20.5\% and stayed constant. Whereas in the container, it started around 1\% and stayed constant there. \emph{Fig: 14 UI JS Service - Load state Comparison} shows the comparison of the CPU utilization and memory consumption by the application on VM and container in the load state. When in the load state the CPU usage by the VM fluctuated at around 25\%, whereas in the container, it fluctuated at around 5\%. The Memory usage by the UI service in VM started around 22.53\% and increased to around 22.67\% during the load, whereas in the container, it is around 1.8\% during the load. By the above two comparisons, we can say that the UI JS service consumes less resources in container than VM, both in idle and load states. \emph{Fig: 15 UI JS Service - Response Time Comparison} shows the comparison of the response times of the UI JS Service when running on VM and Container. While running the application in both VM and containers, 10000 requests were sent to check the response times and the average time was about 376ms when running in the VM, whereas while running in the container the average response time was 140ms. By the above comparison, we can clearly see that the UI JS Service performs better when running on a container.

\paragraph{ Books - Java Service} The Books Micro service was written in Java for the purpose of management of books in the application.  
\begin{figure}[!h]
    \centering
    \includegraphics[width=9.3cm, height=4.8cm]{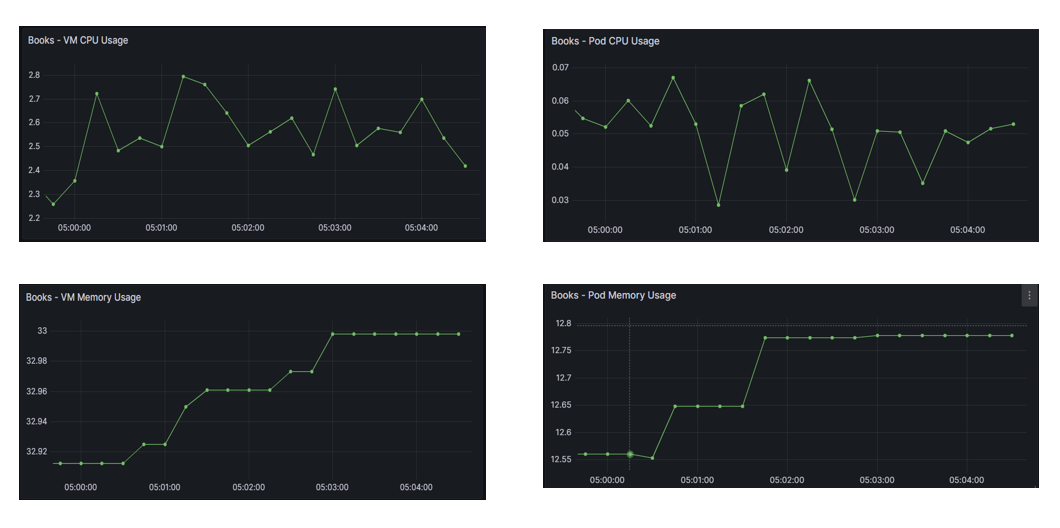}
    \caption{Books Java Service - Idle state Comparison}
\end{figure}
\begin{figure}[!h]
    \centering
    \includegraphics[width=9.3cm, height=4.8cm]{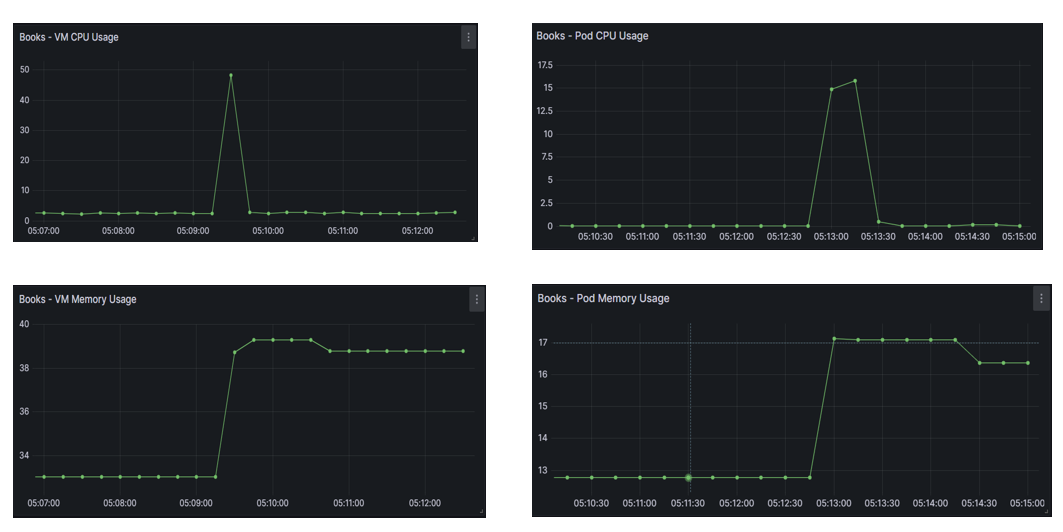}
    \caption{Books Java Service - Load state Comparison}
\end{figure}
\begin{figure}[!h]
    \centering
    \includegraphics[width=9.3cm, height=3cm]{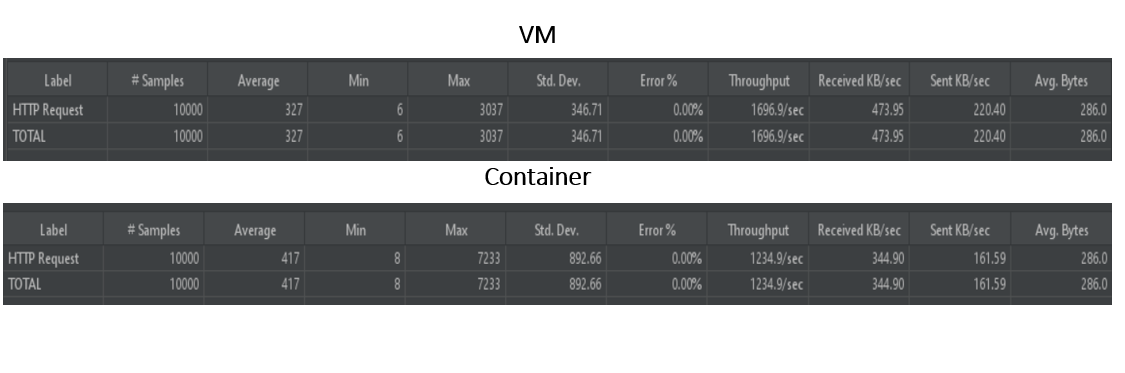}
    \caption{Books Java Service - Response Time Comparison}
\end{figure}
\emph{Fig: 15 Books Java Service - Idle state Comparison} shows the comparison of the CPU utilization and memory consumption by the application on VM and container in the idle state.
CPU Usage: When in idle state the CPU usage by the VM fluctuated between 2.2\% and 2.8\%,whereas in the container, it fluctuated between 0.03\% and around 0.07\%. The memory usage by the order service in VM is around 33\% in the VM whereas in the container, it started around 12.77\%.\emph{Fig: 16 Books Java Service - Load state Comparison} shows the comparison of the CPU utilization and memory consumption by the application on VM and container in the load state. When in load state the CPU usage of the VM is around 50\%, we can see a spike that ranges between 0 and around 50\%. Whereas in the container, it is around 17.5\%. The memory usage by the UI service in VM started is around 32\% and increased to around 40\% during the load, whereas in the container, it started around 13\% and  increased to 17\% during the load.
By the above two comparisons, we can say that the Java service runs better in container than VM, both in idle and load states.\emph{Fig: 17 Books Java Service - Response Time Comparison} shows the comparison of the response times of the Java Service when running on VM and Container.
While running the application in VM and container, 10000 requests were sent to check the response times the average time was about 327ms, whereas while running in the Container the average response time was 417ms. By the above comparison, we can't  actually say that the Java service runs efficiently when ran in a container. Because the response time is more in case of containers. 

\section{Conclusion}
With this evaluation , services running on containers offer several advantages, including faster boot time, lower disk usage, and lower CPU and memory usage. They provide a lightweight and efficient solution for software development and deployment, particularly in resource-constrained environments.
However, Virtual machines take precedence in the situations where the response time is crucial and strict isolation is expected. Container runs as a process on the host machine and provides only isolation at the application level. If security is the major concern, VM offers great level of isolation when compared to containers.

\bibliographystyle{unsrt}
\bibliography{Compiler}
\end{document}